\documentclass[pra,twocolumn,superscriptaddress]{revtex4-2} 
\usepackage{braket,amsmath,amssymb}
\usepackage{mathtools}
\usepackage{mathbbol}
\usepackage{float}
\usepackage{tikz}
\usepackage{verbatim}
\usetikzlibrary{matrix,decorations.pathreplacing, calc, positioning}
\usepackage{color}
\usepackage{enumitem}
\usepackage{graphicx}
\usepackage{dcolumn}
\usepackage{bm}
\usepackage{amsfonts}
\usepackage{mathrsfs}
\usepackage{color,hyperref}
\usepackage{amsmath,tikz-cd} 
\definecolor{darkblue}{RGB}{0,0,139}
\hypersetup{colorlinks,breaklinks,
linkcolor=darkblue,urlcolor=darkblue,
anchorcolor=darkblue,citecolor=
darkblue,pdfauthor=MoChShAr, pdftitle=MoChShAr.Criteria}
\usepackage[normalem]{ulem}

\newcommand{\ie}{\textit{i}.\textit{e}.}

\begin{document}
\title{ 
Continuous variable quantum teleportation,
$U(2)$ invariant squeezing and non-Gaussian resource states}
\author{Mohak Sharma}
\email{ph21012@iisermohali.ac.in}
\affiliation{Department of Physical Sciences,
Indian Institute of Science Education and Research Mohali,
Sector 81 SAS Nagar, Punjab 140306 India.}
\author{Chandan Kumar}
\email{chandan.quantum@gmail.com}
\altaffiliation[Present Address~:~]{ Optics and Quantum
Information Group, The Institute of Mathematical Sciences,
CIT Campus, Taramani, Chennai 600113, India; Homi Bhabha
National Institute, Training School Complex, Anushakti
Nagar, Mumbai 400085, India.}
\affiliation{Department of Physical Sciences,
Indian Institute of Science Education and
Research Mohali, Sector 81 SAS Nagar,
Punjab 140306 India.}
\author{Shikhar Arora}
\email{shikhar.quantum@gmail.com}
\affiliation{Department of Physical Sciences,
Indian Institute of Science Education and
Research Mohali, Sector 81 SAS Nagar,
Punjab 140306 India.}
\author{Arvind}
\email{arvind@iisermohali.ac.in}
\affiliation{Department of Physical Sciences,
Indian
Institute of Science Education and
Research Mohali, Sector 81 SAS Nagar,
Punjab 140306 India.}

\begin{abstract}
We investigate the role of quadrature squeezing in the
quantum teleportation protocol for coherent states, using
non-Gaussian resource states. For the two-mode systems, the
non-Gaussian resource states that we use are obtained by an
experimentally realizable scheme of photon subtraction,
photon addition, and photon catalysis, on the two-mode
squeezed vacuum, and two-mode 
squeezed thermal states.
We first analyze the non-classical attribute of quadrature
squeezing in these generated non-Gaussian states using the
$U(2)$ invariant squeezing approach, which allows us to
account for all possible quadratures. We then show that
the presence of such non-classicality in non-Gaussian
resource states is not necessary for successful quantum teleportation, a finding which is at variance with an earlier result in this direction. This result is important since it demonstrates how non-classicality other than
quadrature squeezing present in the resource can be utilized for quantum teleportation.
\end{abstract} 
\maketitle

%

\section{Introduction}
Quantum teleportation is a concept in quantum information
science that allows for the sharing of information using
shared entanglement and has applications in various fields
including long-distance communication, distributed quantum
computing, and remote state
preparation~\cite{Pirandola2015}. It was first conceived for
discrete variable (DV) systems~\cite{bennet} and
subsequently extended to continuous variable (CV)
systems~\cite{bk-1998}. The theoretical proposal was
followed by experimental realizations of
DV~\cite{ExperimentalDV,Zeilinger,Zeilinger2004} and CV
teleportation~\cite{science,Kimble,cvexperimental}. Recent
studies have utilized non-Gaussian operations to enhance
the performance of various CV quantum information processing
tasks such as quantum
metrology~\cite{gerryc-pra-2012,josab-2012,braun-pra-2014,josab-2016,pra-catalysis-2021,crs-ngtmsv-met,ngsvs,metro-thermal},
quantum
illumination~\cite{ill2008,ill2013,illumination14,illumination18,rivu}
and quantum
teleportation~\cite{tel2000,Akira-pra-2006,Anno-2007,tel2009,wang2015,catalysis15,catalysis17,tele-arxiv,better}.

While entanglement is a prerequisite for quantum
teleportation~\cite{Braunstein-jmo-2000,Braunstein-pra-2001,Pirandola2006},
a significant amount of effort has been put into identifying
other quantum attributes that can enhance and assist in
quantum teleportation. For instance, Ref.~\cite{dellano2007}
demonstrated that teleportation fidelity depends on
entanglement, non-Gaussianity and squeezed vacuum affinity
in a nontrivial manner. The role of Einstein-Podolsky-Rosen (EPR) correlations in quantum
teleportation has been explored in
detail~\cite{lee2011,lee2013,catalysis15,catalysis18}.
Examples of resource states, which do not exhibit
second-order EPR correlation, but exhibit the fidelity of
teleportation of a coherent state exceeding 1/2 (indicating
success of quantum
teleportation~\cite{Braunstein-jmo-2000,Braunstein-pra-2001}),
were found in Refs.~\cite{lee2013,yang2009}. Similarly, examples
of states with EPR correlation that do not yield quantum
teleportation were also found in Refs.~\cite{lee2011,lee2013}.
Along the same lines, we have performed an analysis for
exploring the intersection between successful quantum
teleportation, and quadrature squeezing, which is a well
known signature of non-classicality. We have generated
non-Gaussian states using practical schemes of photon
subtraction (PS), photon addition (PA), and photon
catalysis (PC) on Gaussian states such as 
the two-mode squeezed
vacuum (TMSV) and the two-mode squeezed thermal (TMST)
state~\cite{Bartley_2015,tele-arxiv}.

How do these states perform while carrying out quantum
teleportation in different parameter ranges is the
subject of the current study. In  particular, we want to find out if teleportation
can be carried out with states in this family when there is
no quadrature squeezing? We find that, for both pure and
mixed non-Gaussian resource states that we
consider, there exist a parameter regime where successful
teleportation of coherent states is possible without the
resource state possessing quadrature squeezing.

This result, therefore,
unpins the teleportation success from the presence of
quadrature squeezing, and is in contrast with the results
of~\cite{Bose}, where it is argued that quadrature squeezing
appears to be essential for success of quantum teleportation
even with pure non-Gaussian resource states.  To note, by
successful quantum teleportation, here it is meant that the
fidelity of teleporting an unknown coherent state from a
uniform distribution of coherent states is greater than
$1/2$, which is the upper bound of fidelity achievable
through any classical
strategy~\cite{Braunstein-jmo-2000,Braunstein-pra-2001}.

It is also important to note
that for Gaussian states it
is not possible to have non-classicality and therefore
entanglement without quadrature squeezing~\cite{Zubairy15}.
The non-Gaussian states, on the other hand, are not bound by
this restriction, as demonstrated in this paper, where
specific non-Gaussian states exhibiting entanglement - a
hallmark of non-classicality- are shown to lack quadrature
squeezing.  That such non-classicality in non-Gaussian
states without quadrature squeezing can actually be used for
quantum teleportation is demonstrated in this paper.  Since
squeezing can be present in any of the quadratures of the
two-mode system, the $U(2)$ invariant squeezing approach is useful for our
analysis~\cite{simon-1994,arvind_pra_1995,ARVIND2002461}.

The article is arranged as follows: We begin by 
recapitulating the necessity of quadrature squeezing in
Gaussian resource based CV quantum teleportation in
Sec.~\ref{gqt}. In Sec.~\ref{ngqt}, we consider both pure
and mixed non-Gaussian resource states and analyse the
teleportation fidelity along with quadrature squeezing in the
parameter space. To this end, we find out whether quadrature
squeezing is a necessary condition for successful CV quantum
teleportation for these non-Gaussian resource states. The conclusions 
of our study are presented
in Sec.~\ref{sec:conc}.
Additionally, we have added an appendix which covers an
overview of CV systems, and some essential prerequisites
necessary to follow the current article.
\section{Role of $U(2)$ invariant quadrature squeezing in
Gaussian and non-Gaussian CV quantum teleportation}
\subsection{The case of Gaussian resource states}
\label{gqt}
According to $U(2)$ invariant
squeezing criterion, a state is said to be squeezed if and
only if the least eigenvalue($\lambda_{\text{min}}$) of the
corresponding covariance matrix is less than
$1/2$~\cite{simon-1994,arvind_pra_1995,ARVIND2002461}. The
necessity of $U(2)$ invariant quadrature squeezing for CV
quantum teleportation using two-mode Gaussian
resource states is evident from the following argument. It
has been shown that if the covariance matrix $\sigma$ of a
two-mode Gaussian state satisfies $\sigma \geq \mathbb{1}/2$, the
state is deemed classical~\cite{Simonprl2000,Serafini}, in
the sense that its $P$-function is a positive valued proper
function~\cite{Sudarshan,Glauber}. Such a condition on its
$P$-function will imply the state to be separable~\cite{Lee,
Serafini}. Thus, from the positive semi-definiteness of the matrix $\sigma-\frac{\mathbb{1}}{2}$, we have,
 
\begin{equation}
\lambda_{\text{min}} \geq \frac{1}{2} \Rightarrow \text{ state is separable}.
\end{equation}
The contrapositive of the above implication would be,
\begin{equation}
\text{ state is inseparable} 
\Rightarrow  \lambda_{\text{min}} < 1/2.
\end{equation}
Thus, an inseparable (entangled) state implies quadrature
squeezing \ie, quadrature squeezing is necessary for
entanglement and, consequently, for quantum
teleportation~\cite{Pirandola2006, Braunstein2001pra,
Samuelpra}. Quadrature squeezing, clearly is not sufficient
as there are quadrature squeezed states which do not possess
any entanglement and they obviously cannot be used for
quantum teleportation.
\subsection{The case of non-Gaussian resource states}
\label{ngqt}
\begin{figure}[h!]
\includegraphics[scale=1]{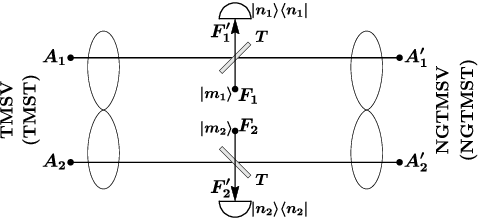} \caption{
Practical setup for the implementation of non-Gaussian
operations on TMSV (or TMST) state. We start by combining
modes $A_1$ and $A_2$ of the TMSV (or TMST) state with modes
$F_1$ and $F_2$, which are prepared in the Fock states
$|m_1\rangle$ and $|m_2\rangle$, respectively. This is done
using beam splitters of transmissivity $T$. Detection of
$n_1$ and $n_2$ photons in modes $F'_1$ and $F'_2$
indicates successful implementation of non-Gaussian
operations on the modes $A_1$ and $A_2$ of the TMSV state,
respectively. } \label{figtmsv} \end{figure}

In this section we consider pure and mixed non-Gaussian
resource states for teleportation of input coherent states.
In contrast to the claim made in Ref.~\cite{Bose}, we show
that for both pure and mixed cases, quantum teleportation is
possible without quadrature squeezing. The schematic
for generating the non-Gaussians states from two mode
squeezed vacuum and from two mode squeezed thermal states is
depicted in Fig.~\ref{figtmsv}.
\subsubsection{Teleportation using a class of pure non-Gaussian
 resource states}
\label{sec:char}
We first  consider pure
non-Gaussian resource states generated by performing
non-Gaussian operations, namely, PS, PA, and PC on the TMSV
state for the teleportation of a coherent state. As shown in
Fig.~\ref{figtmsv}, the modes
$A_1$ and $A_2$ of the TMSV state are combined  with
Fock states $|m_1\rangle$ and $|m_2\rangle$ using beam
splitters. The output auxiliary modes are then subjected to
photon number resolving measurements and the detection of
$|n_1\rangle$ and $|n_2\rangle$ photons signals successful
non-Gaussian operations on both the modes of the TMSV state. The three distinct non-Gaussian
operations on mode $A_i$ can be envisaged depending on
the relative magnitudes of $m_i$ and $n_i$. The conditions
$m_i<n_i$, $m_i>n_i$, and $m_i=n_i$ correspond to PS, PA,
and PC operations, respectively. The actions of these
operations on TMSV state lead to the generation of PSTMSV,
PATMSV, and PCTMSV states, respectively. We will be
representing the generated non-Gaussian (NG) states in the
following form: $(m_1, n_1)~(m_2 ,n_2)$-NGTMSV. We have
restricted our analysis to symmetric single photon
subtraction, addition and catalysis operations.

We employ the Wigner characteristic function formalism for
the purpose of our calculations. The notations and
symbols used in the following calculations, unless
explicitly stated, are defined in the appendix. Our aim is
to first determine the characteristic functions for our
generated non-Gaussian states, from which we will obtain the
teleportation fidelity and the associated squeezing
properties of the states. We start with writing the wigner
characteristic function of the TMSV state characterized by a
finite squeezing parameter $r$ (See Eq
(\ref{gaussianwigchar}) and  Sec.~\ref{tmsvtmstbrief} for
details):
\begin{equation}
\begin{aligned}
\chi_{\text{TMSV}}(\Lambda_1, \Lambda_2)= & \exp
\bigg[-\frac{1}{4} \cosh (2
r)(\tau_1^2+\sigma_1^2+\tau_2^2+\sigma_2^2) \\
&\quad\quad\quad +\frac{\tau_1 \tau_2-\sigma_1 \sigma_2}{2} \sinh
(2 r)] \bigg].
\end{aligned}
\end{equation}
Here, and in the subsequent parts of the article,
$\Lambda_i= (\tau_i, \sigma_i)$. Prior to the beam
splitter operations, the Wigner characteristic function of
the joint state of the system consisting of the TMSV state
and the ancilla modes can be written as,
\begin{equation}
\chi_{F_1 A_1 A_2
F_2}(\Lambda^{\prime})=\chi_{\text{TMSV}}\left(\Lambda_1,
\Lambda_2\right) \chi_{|m_{1}\rangle}\left(\Lambda_3\right) 
\chi_{|m_{2}\rangle}\left(\Lambda_4\right),
\end{equation}
where $\Lambda^{\prime} = \left(\Lambda_1, \Lambda_3,
\Lambda_2, \Lambda_4\right)^{T}$. \\

Post the beam splitter operations, the Wigner characteristic
function of the joint state can be written using the
transformation rule given in Eq.~(\ref{transform}) as 
\begin{equation}
\chi_{F_1' A_1' A_2' F_2'}(\Lambda^{\prime})=\chi_{F_1 A_1
A_2 F_2}\left([B_{13}(T)\oplus B_{24}(T)]^{-1}
\Lambda^{\prime}\right).
\end{equation}

Finally, upon the required photon number resolving
measurements on the output auxiliary modes, the
characteristic function of the unnormalized final state
turns out to be~\cite{olivares-2012},
\begin{equation}\label{detect}
\begin{aligned}
\chi^{\prime}_{\text{NGTMSV}}(\Lambda_1, \Lambda_2)=
\frac{1}{(2
\pi)^2} \int & d^2 \Lambda_3 d^2 \Lambda_4 
\chi_{F_1' A_1' A_2'
F_2'}(\Lambda^{\prime})\\
&\times 
\chi_{|n_1\rangle
}(\Lambda_3)
\chi_{|n_2\rangle
}(\Lambda_4) \\
\end{aligned}
\end{equation}
The normalized Wigner characteristic function is given by,
\begin{equation}
\chi_{\text{NGTMSV}}\left(\tau_1, \sigma_1, \tau_2,
\sigma_2\right)= \frac{1}{P_\text{ NG}}
\chi_{\text{NGTMSV}}^{\prime}\left(\tau_1, \sigma_1, \tau_2,
\sigma_2\right),
\end{equation}
where $P_{\text{ NG}}$ is the success probability of the
non-Gaussian operation obtained by,
\begin{equation}
P_{\text{ NG}}=\chi_{\text{NGTMSV}}^{\prime}\left(\tau_1,
\sigma_1, \tau_2, \sigma_2\right)
|_{\tau_1=\sigma_1=\tau_2=\sigma_2=0}.\\
\end{equation}
For the sake of completeness, we mention that in the limit
of $T\rightarrow 1$, the experimental scheme considered for
photon subtraction and addition, reduces to the ideal photon
subtraction ($\hat{a}$) and addition ($\hat{a}^{\dagger}$)
operations, while that corresponding to catalysis reduces to
the identity operation. In this limit, the success
probability $P_{\text{ NG}}$ approaches zero.

For obtaining the fidelity of teleportation of coherent
states with the generated non-Gaussian resource states, we
employ Eq.(\ref{fidex1}) and Eq.(\ref{teleported}), allowing
us to compute it from the characteristic functions of the
input coherent state and the non-Gaussian resource state.
The analytical results for the PSTMSV and PATMSV cases have
been mentioned in Table~\ref{table1}. 

\begin{figure*} 
\begin{center}
\includegraphics[scale=1]{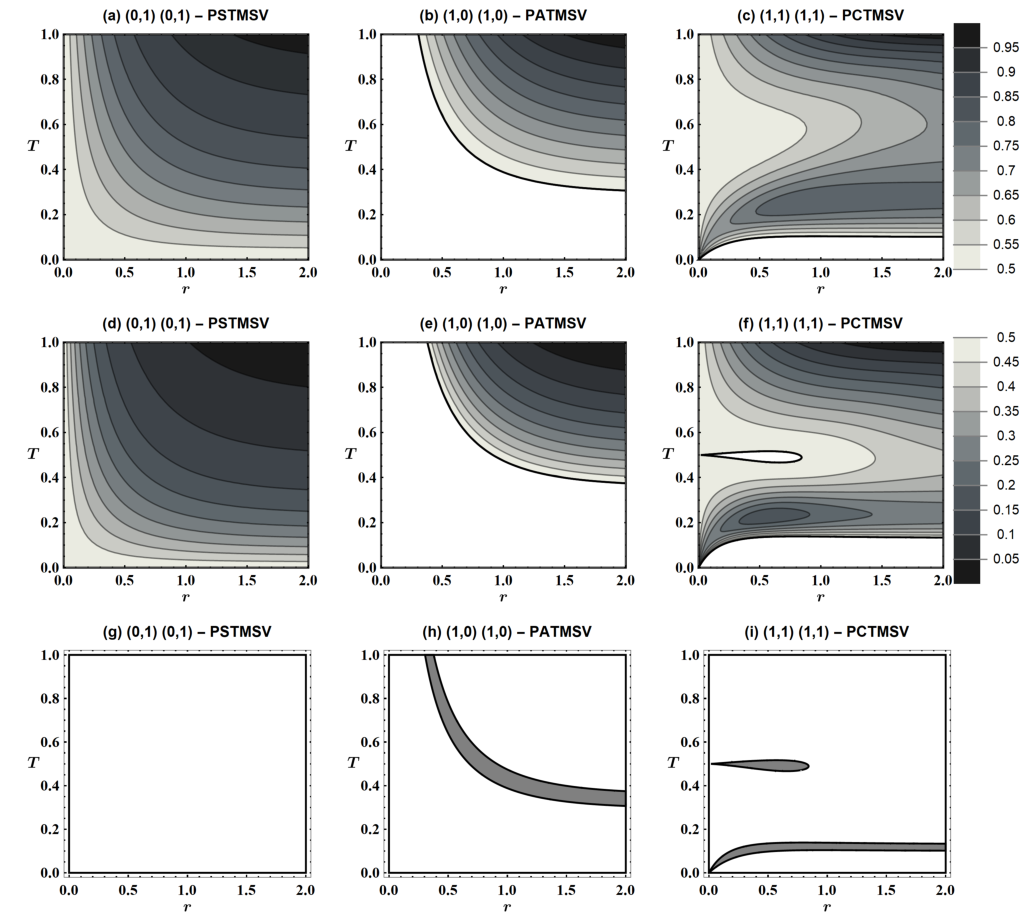}
\caption{Top Row ((a)-(c)): Contour plots for the NGTMSV states depicting the
variation of fidelity $F$ in the $(r, T)$ parameter space, Middle Row ((d)-(f)): Contour plots for the NGTMSV states depicting the variation of minimum eigenvalue $\lambda_{\text{min}}$ in the $(r, T)$ parameter space, Bottom Row ((g)-(i)): Region plots for the NGTMSV states  in $(r, T)$ parameter space, depicting the regions where successful quantum teleportation is possible in the absence of quadrature squeezing.
 }
\label{paneltmsv}
\end{center}
\end{figure*}

The analytical  expression  of fidelity for the catalysis case
turns out to be quite cumbersome and therefore we have not
displayed it.  All fidelity results are shown as contour
plots in the $(r, T)$ parameter space which provides an
intuitive visualization of the varying fidelity value. The
plots are displayed in Fig.~\ref{paneltmsv}(a)-(c)~(top
row).The shaded regions represent resource states
corresponding to which the teleportation fidelity $F>1/2$,
while the unshaded white regions along with the black
boundary represent states corresponding to which $F \leq
1/2$, with the equality holding up on the black boundary.

We now move on to analyse the $U(2)$ invariant squeezing
present in the non-Gaussian states generated in the
processes of PS, PA and PC. To this end, we compute their
respective covariance matrices and their  minimum
eigenvalues $(\lambda_{\text{min}})$ and use this minimum
eigenvalue as the quantifier of the $U(2)$ invariant
quadrature squeezing present in the state. Quite remarkably,
the elements of the covariance matrix corresponding to a
state can be easily obtained from its Wigner characteristic
function. The following identity, which can be used to
obtain the average of symmetrically ordered operators of the
form $\{\hat{q}_1^{r_1}
\hat{p}_1^{s_1}{\hat{q_2}}^{r_2}{\hat{p_2}}^{s_2}\}_{\mathrm{sym}}$,
serves useful for this purpose~\cite{olivares-2012}:
\begin{equation}\label{app:covfinalch}
\begin{aligned}
\langle\{\hat{q}_1^{r_1}
\hat{p}_1^{s_1}{\hat{q_2}}^{r_2}{\hat{p_2}}^{s_2}\}_{\mathrm{sym}}\rangle
=
\widehat{O} \chi_{\text{NGTMSV}} (\tau_1,
\sigma_1,\tau_2, \sigma_2) , 
\end{aligned}
\end{equation}
where 
\begin{equation}
\begin{aligned}
\widehat{O} =&\left( \frac{1}{i} \right)^{r_1+r_2}\left(
\frac{1}{-i} \right)^{s_1+s_2}
\frac{\partial^{r_1+s_1}}{\partial \sigma_1^{r_1} \partial
\tau_1^{s_1} }\\
&\times\frac{\partial^{r_2+s_2}}{\partial \sigma_2^{r_2}
\partial \tau_2^{s_2} } \{ \bullet \}_{\substack{\tau_1=
\sigma_1=0\\ \tau_2=
\sigma_2=0}},
\end{aligned}
\end{equation}
\begin{center}
\begin{table}[ht!]
\caption{\label{table1}
Fidelity and minimum eigenvalue expressions for
$(0,1)~(0,1)$-PSTMSV and $(1,0)~(1,0)$-PATMSV states. Here,
$\alpha=T^2 \tanh{r}$, and $\beta=T \tanh{r}$.}
\begin{tabular}{|| l | c | c || r }
\hline \hline
~ &~&~\\
&
$(0,1)(0,1)$-PSTMSV & $(1,0)(1,0)$-PATMSV \\
~ &~&~\\
\hline \hline
~ &~&~\\
$~F$& $\dfrac{(1+\alpha)^3 }{4 \left(1+\alpha
^2\right)}\left(1+(1-\alpha )^2\right)$&
$\dfrac{(1+\alpha)^3}{4 \left(1+\alpha ^2\right)}$ \\ ~ &~&~\\
\hline ~ &~&~\\ $\lambda_{\text{min}}$&$\dfrac{(1-\beta )^2
}{2 \left(1-\beta ^4\right)}\left(1-2 \beta +3 \beta
^2\right)$& $\dfrac{(1-\beta )^2}{2 \left(1-\beta
^4\right)}\left(3-2 \beta +\beta ^2\right)$\\ ~ &~&~\\
\hline \hline
\end{tabular} \end{table} \end{center}

We have presented the results of the variation in the
minimum eigenvalue for the non-Gaussian resource states in
Fig.~\ref{paneltmsv}~(d)-(f)~(middle row), which shows contour
plots in the
$(r,T)$ parameter space. The shaded regions are the ones
that correspond to the minimum eigenvalue of the covariance
matrix being less than one-half $(\lambda_{\text{min}}<
1/2)$, and are said to represent states exhibiting the
non-classical signature of quadrature squeezing. The
unshaded white regions along with the black
boundary in the plots denote non-Gaussian states
that are devoid of quadrature squeezing
$(\lambda_{\text{min}} \geq 1/2)$. Specifically,
$\lambda_{\text{min}} = 1/2$ for the states characterized by
the black boundary,  while for the unshaded regions,
$\lambda_{\text{min}} > 1/2)$. 
Interestingly, we
observe that the
application of PA and PC operations on the TMSV state has
the potential to kill the squeezing in the non-Gaussian state, which was present in the initial
TMSV state (the TMSV state being quadrature squeezed for all values of squeezing parameter $r$). We have displayed the explicit expressions for
the minimum eigenvalue corresponding to PSTMSV and PATMSV
states in Table~\ref{table1}. Again, we have omitted
mentioning the explicit expression for the PCTMSV case due
to its complicated form.


To view if successful quantum teleportation is possible
without quadrature squeezing, we display the regions of $(r,
T)$ parameter space corresponding to a fidelity value
greater than half as well as the  minimum eigenvalue greater
than half in Fig.~\ref{paneltmsv} (bottom row).  As can be
seen for the case of PSTMSV (Fig.~\ref{paneltmsv}(g)), we do
not observe any shaded region, implying
that there are no states which
can be used for teleportation, and which are without
quadrature squeezing.  Moving on to the PATMSV
(Fig.~\ref{paneltmsv}(h))  and PCTMSV states
(Fig.~\ref{paneltmsv}(i)), we clearly find regions in the
$(r, T)$ parameter space where we do have teleportation
fidelity greater than half and minimum eigenvalue also
greater than or equal to half.
Hence, such states are our
desired states which can lead to successful quantum
teleportation in the absence of quadrature squeezing.

These results are therefore at variance with the proposition
stated in Ref.~\cite{Bose}, wherein it is mentioned that
quadrature squeezing in non-Gaussian resource states appears
to be a necessary condition for successful CV quantum
teleportation of coherent states. We further observe that,
although quadrature squeezing in non-Gaussian states is no
longer an essential requirement for successful CV quantum
teleportation, for the specific class of non-Gaussian states
that we have considered, the presence of strong squeezing
notably appears to correlate with a high teleportation
fidelity value. This is very likely due to the fact that, in
our original Gaussian states, which we modified to obtain
the non-Gaussian states, the source of non-classicality was
quadrature squeezing.
\subsubsection{Teleportation using a class of mixed non-Gaussian
resource states}
\begin{figure*}
\begin{center}
\includegraphics[scale=1]{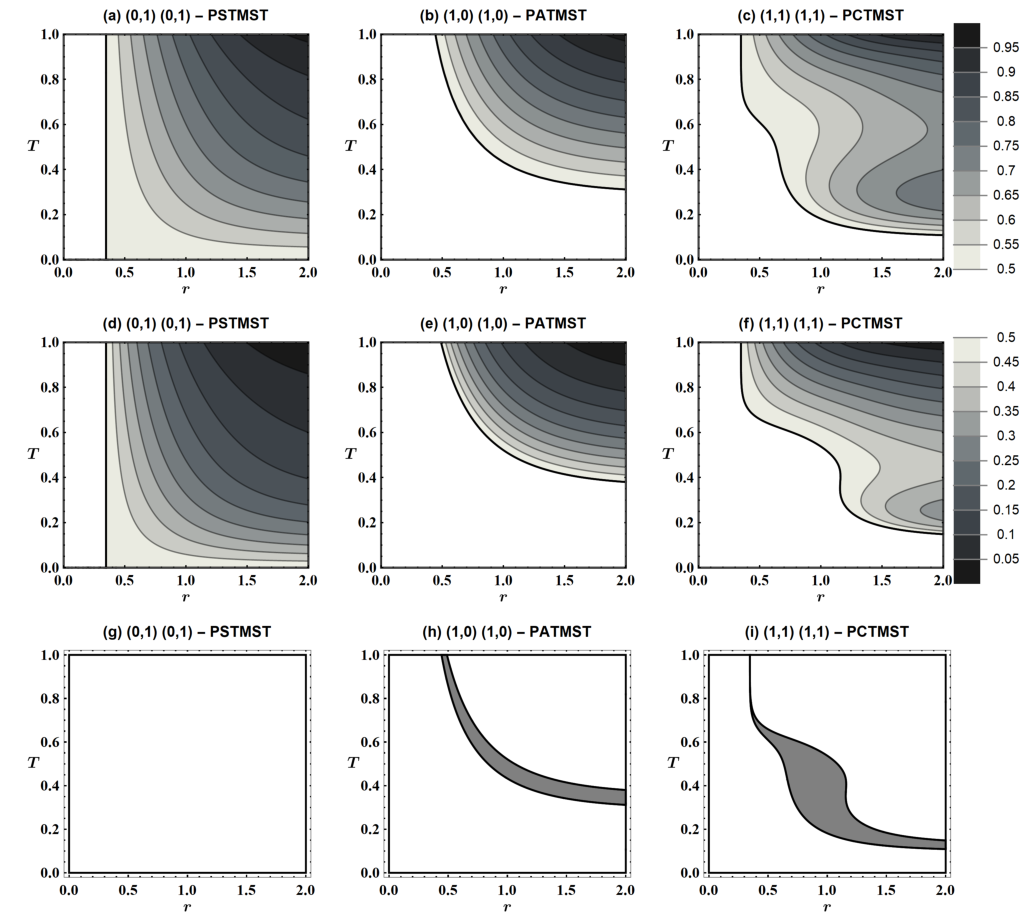}
\caption{Top Row ((a)-(c)): Contour plots for the
NGTMST states depicting the variation of fidelity $F$ in the
$(r, T)$ parameter space, Middle Row ((d)-(f)): Contour
plots for the NGTMST states depicting the variation of
minimum eigenvalue $\lambda_{\text{min}}$ in the $(r, T)$
parameter space, Bottom Row ((g)-(i)): Region plots for the
NGTMST states  in $(r, T)$ parameter space, depicting the
regions where successful quantum teleportation is possible
in the absence of quadrature squeezing.}
\label{paneltmst}
\end{center}
\end{figure*}
We now turn to the consideration of resource states
generated by performing non-Gaussian operations (PS, PA and
PC) on a TMST state. The TMST state is a well-known example
of a Gaussian mixed state and has been generated in the
laboratory~\cite{tmst-prl-2014,tmst-prl-2016,tmst-2016,tmst-njp-2016}.

We adopt the same procedure as described in
Sec.~\ref{sec:char} for implementing the non-Gaussian (NG)
operations on the TMST state, as illustrated in
Fig.~\ref{figtmsv}, to generate the non-Gaussian states.
But this time, we use the TMST state instead of the
TMSV state as the input. We represent the generated state
by: $(m_1, n_1)~(m_2 ,n_2)$-NGTMST. We have chosen
$\kappa=1$ for the purpose of analysis in the current study
(see Eq.~\ref{covmatrixTMST}). The same methodology as
described in Sec.~\ref{ngqt}~(1) is be used to obtain the
fidelity of teleportation and the minimum eigenvalue of the
covariance matrix, the results of which are displayed as
contours in Fig.~\ref{paneltmst}~(a)-(c) (top row) and
Fig.~\ref{paneltmst}~(d)-(f) (middle row) respectively.
These contour plots in the $(r, T)$ parameter space,
respectively, quantify the success in CV quantum
teleportation, and the quadrature squeezing present in the
generated mixed non-Gaussian states. In
Fig.~\ref{paneltmst}~(a)-(c), the shaded regions depict the
regions where teleportation fidelity $F>1/2$, while the
unshaded regions correspond to $F<1/2$. The black boundary
separating the shaded and unshaded regions represents
resource states for which $F=1/2$. Moving on to
Fig.~\ref{paneltmst}~(d)-(f), the shaded regions here
correspond to the existence of quadrature squeezing in the
resource $(\lambda_{\text{min}} < 1/2)$, while the unshaded
regions along with the black boundary represent states which
are not quadrature squeezed $(\lambda_{\text{min}} \geq
1/2)$, with the equality holding on the black boundary. The
analytical expressions for the cases have been avoided here
due to their complexity.

Similar to the previous section, to view the relevance of
quadrature squeezing for quantum teleportation, we have
shaded those regions in the parameter space for which the
corresponding fidelity value is greater than half ($F>1/2$)
along with the corresponding states not being quadrature
squeezed ($\lambda_{\text{min}} \geq 1/2$). The results for
the cases are shown in Fig.~\ref{paneltmst} (bottom row).
For the case describing PSTMST (Fig.~\ref{paneltmst}(g)), we
observe no such states. In contrast, both PATMST
(Fig.~\ref{paneltmst}(h)) and PCTMST states
(Fig.~\ref{paneltmst}(i)) have such regions thereby implying
the possibility of such states. Similar to the results in
the previous section, for the particular class of NGTMST
states that we have considered, we find that though
quadrature squeezing in the resource is not a necessary
requirement for exhibiting successful CV quantum
teleportation, still it so occurs that the states leading to
a high fidelity of teleportation seem to display a
correspondence with strong squeezing. Secondly, the region in
the parameter space where successful teleportation is
possible without quadrature squeezing is larger for the case
when we start with TMST states as compared that for TMSV
state.

\section{Conclusions}
\label{sec:conc}
In this study, we investigated the importance of $U(2)$
invariant squeezing in the context of continuous variable
quantum teleportation. We first reviewed the well-known case
of Gaussian resources, for which the necessity of quadrature
squeezing for CV quantum teleportation can be seen in a
straightforward way. We then
moved on to the analysis of quantum teleportation using
non-Gaussian resource states. Specifically, we considered
the non-Gaussian states generated by experimentally feasible
photon subtraction, addition and catalysis operations on the
TMSV and TMST states.  The former case led to generation of
pure non-Gaussian states, while the latter led to mixed
non-Gaussian states. The teleportation fidelity and
squeezing properties were calculated for these classes of
non-Gaussian states and the results were plotted as
parametric plots to identify regions of successful
teleportation and their correlations with quadrature
squeezing. This aided in identifying the hitherto
unknown regions in the parameter space where successful
teleportation can be carried out without the presence of
quadrature squeezing.

Our analysis is in contrast to the conclusions made
in~\cite{Bose} where it is asserted that quadrature
squeezing is necessary for successful quantum teleportation
even for non-Gaussian resource states. Finally, while quadrature squeezing in Gaussian
resource states is a necessary condition for successful CV
quantum teleportation, we can have non-Gaussian states with
no quadrature squeezing which can be successfully used as
resource states for quantum teleportation.

\appendix
\section{A Brief Review of CV Systems and VBK protocol for
Quantum Teleportation}
\subsection{Review of CV Systems}
\label{basicreviewCV}
An $n$-mode CV system can be described by $n$ pairs of
Hermitian quadrature operators $\hat{q}_i,\hat{p}_i$
($i=1\,,2,\dots,
n$)~\cite{arvind1995,Braunstein,adesso-2007,weedbrook-rmp-2012,adesso-2014}.
These quadrature operators can be arranged as a column
vector:
\begin{equation}\label{eq:columreal}
\hat{ \xi} =(\hat{ \xi}_i)= (\hat{q_{1}},\,
\hat{p_{1}} \dots, \hat{q_{n}}, 
\, \hat{p_{n}})^{T}, \quad i = 1,2, \dots ,2n,
\end{equation}
which allows us to express the bosonic commutation relation
compactly as 
\begin{equation}\label{eq:ccr}
[\hat{\xi}_i, \hat{\xi}_j] = i \Omega_{ij}, \quad (i,j=1,2,...,2n).
\end{equation}
Here, we have set $\hbar$=1 and 
$\Omega$ is the symplectic form, which can be written as
\begin{equation}
\Omega = \bigoplus_{k=1}^{n}\omega = \begin{pmatrix}
\omega & & \\
& \ddots& \\
& & \omega
\end{pmatrix}, \quad \omega = \begin{pmatrix}
0& 1\\
-1&0 
\end{pmatrix}.
\end{equation}
An $n$-mode CV system can also be described by 
$n$-pairs of annihilation $\hat{a}_i$ and creation operators 
$\hat{a}^{\dagger}_i$ $(i=1,2,...,n)$. These are related to the 
quadrature operators as follows:
\begin{equation}\label{realtocom}
\hat{a}_i= \frac{1}{\sqrt{2}}(\hat{q}_i+i\hat{p}_i),
\quad \hat{a}^{\dagger}_i= \frac{1}{\sqrt{2}}(\hat{q}_i-i\hat{p}_i).
\end{equation}
The Fock states $\vert n_i \rangle, \quad \{n_i=0,\,1, \dots
,\infty \} $, the eigenvectors of the number operator
$N_i=a_i^{\dagger} a_i$ with eigenvalues $n_i$, form the
basis of the Hilbert space of the $i^{\text{th}}$ mode,
Similarly, $ \vert n_1\dots n_i \dots n_n\rangle$ with
$\{n_1,\, \dots\,, n_i,\, \dots\,, n_n=0,\, 1, \dots ,\infty
\} $ form the basis of the combined Hilbert space
$\mathcal{H}^{\otimes n} = \otimes_{i=1}^{n}\mathcal{H}_i$
of the $n$-mode system.
\subsubsection{Symplectic operations}
Consider the linear homogeneous transformations represented
by real $2n \times 2n$ matrices $S$ acting on the quadrature
operators as $\hat{\xi}_i \rightarrow \hat{\xi}_i^{\prime} =
S_{ij}\hat{\xi}_{j}$. Requiring that the transformation
should preserve the canonical commutation
relations~(\ref{eq:ccr}) leads to the condition $S\Omega S^T
= \Omega$.  The set of all matrices $S$ satisfying the
aforementioned condition form the symplectic group
$Sp(2n,\,\mathcal{R})$~\cite{arvind_pra_1995}.

Below, we describe the two 
basic symplectic operations that have been used extensively
in the current article.
\\
\par
\noindent
{\bf Single mode squeezing transformation:}
It acts
on the quadrature operators $(\hat{q}_i, \hat{p}_i)^T$ 
via the matrix
\begin{equation}
S_i(r) = \begin{pmatrix}
e^{-r} & 0 \\
0 & e^{r}
\end{pmatrix}.
\end{equation}
The corresponding unitary operation $\mathcal{U}(S_i(r))$ is given by,
\begin{equation}
\mathcal{U}\left(S_i(r)\right)=\exp \left[r\left(a_i^2-\hat{a}_i^{\dagger^2}\right) / 2\right]
\end{equation}
\par
\noindent
{\bf Beam splitter operation\,:}
It acts on the quadrature operators of two-mode systems 
$ \hat{\xi} = (\hat{q}_{i}, \,\hat{p}_{i},\, \hat{q}_{j},\,
\hat{p}_{j})^{T}$ via the matrix
\begin{equation}\label{beamsplitter}
B_{ij}(\theta) = \begin{pmatrix}
\cos \theta \,\mathbb{1}_2& \sin \theta \,\mathbb{1}_2 \\
-\sin \theta \,\mathbb{1}_2& \cos \theta \,\mathbb{1}_2
\end{pmatrix},
\end{equation}
where $\theta$ is related to transmissivity $\tau$ via 
the relation $\tau = \cos ^2 \theta$. Putting $\theta =
\pi/4$ gives $\tau=1/2$, which corresponds to a balanced
(50:50)
beam splitter. 
\\ The corresponding unitary operation $\mathcal{U}(B_{i j}(\theta))$ is given by,
\begin{equation}
\mathcal{U}(B_{i j}(\theta))=\exp [\theta(\hat{a}_i^{\dagger} \hat{a}_j-\hat{a}_i \hat{a}_j^{\dagger})] .
\end{equation}
\subsubsection{Phase space representation via Wigner
characteristic function}
The Wigner characteristic function of a density operator 
$\hat{\rho}$ of an $n$-mode quantum system is given by
\begin{equation}\label{wigdef}
\chi(\Lambda) = \text{Tr}[\hat{\rho} \, \exp(-i \Lambda^T
\Omega \hat{\xi})],
\end{equation}
where $\xi = (\hat{q_1}, \hat{p_1},\dots \hat{q_n}, \hat{p_n})^T$, 
$\Lambda = (\Lambda_1, \Lambda_2, \dots \Lambda_n)^T$ with 
$\Lambda_i = (\tau_i, \sigma_i)^T \in \mathcal{R}^2$.
This function is uniquely determined for a density operator
$\rho$. There is an important class of CV system states,
called Gaussian states, which are completely characterized
by first and second statistical moments. The first
statistical moments can be written in the form of a vector,
\begin{equation}
\bm{d} = \langle \hat{\xi } \rangle =
\text{Tr}[\hat{\rho} \hat{\xi}].
\end{equation}
The second statistical moments can be arranged in the form of a 
real symmetric $2n\times2n$ matrix termed covariance matrix
which is given by
\begin{equation}\label{eq:cov}
V = (V_{ij})=\frac{1}{2}\langle \{\Delta \hat{\xi}_i,\Delta
\hat{\xi}_j\} \rangle,
\end{equation}
where $\Delta \hat{\xi}_i = \hat{\xi}_i-\langle \hat{\xi}_i
\rangle$, and $\{\,, \, \}$ denotes anti-commutator.
For a physically realizable state, $V+i\Omega/2 \geq 0$. The
characteristic function of a Gaussian state can be obtained
through these moments by,
\begin{equation}
\label{gaussianwigchar}
\chi(\Lambda)=\exp \left[-\frac{1}{2} \Lambda^T\left(\Omega
V \Omega^T\right) \Lambda-i(\Omega d)^T \Lambda\right]
\end{equation}

Under a symplectic transformation $S$, the mean vector
$\bm{d}$, covariance matrix
$V$ and Wigner characteristic functions $\chi(\Lambda)$
transform as~\cite{Serafini},
\begin{equation}
\label{transform}
\bm{d} \rightarrow S \bm{d},  V \rightarrow SVS^{T},  \chi(\Lambda)
\rightarrow \chi(S^{-1}\Lambda).
\end{equation}
\subsubsection{Brief Description of TMSV and TMST states}
\label{tmsvtmstbrief}
Below, we give a brief description of the two important
Gaussian states that have been used in the current article:
the two-mode squeezed vacuum (TMSV) state, and the two mode
squeezed thermal (TMST) state. Both TMSV and TMST states are
completely characterized by their mean vector and covariance
matrix as they are Gaussian.

To generate the TMSV state, we begin with the two mode
vacuum state $\bm{|0\rangle}=|0\rangle\otimes|0\rangle$ and
respectively squeeze the two modes by equal and opposite
amounts $r$, i.e. $\mathcal{U}(S_{1}(r))|0\rangle\otimes
\mathcal{U}(S_{2}(-r))|0\rangle$. 
 As per Eq.~(\ref{transform}), the covariance matrix of this
state can be written as,
\begin{equation}
\label{tmsvcov}
\begin{aligned}
V_{\text{initial}}&= [S_1(r)\oplus S_2(-r)][V_{|0\rangle}
\oplus V_{|0\rangle}] [S_1(r)\oplus S_2(-r)]^T, \\
&=\frac{1}{2} \left(
\begin{array}{cccc}
 a^{-2 } & 0 & 0 & 0 \\
 0 & a^{2 } & 0 & 0 \\
 0 & 0 & a^{2 } & 0 \\
 0 & 0 & 0 & a^{-2 } \\
\end{array}
\right),
\end{aligned}
\end{equation}
where $a=e^r$, and $V_{|0\rangle}$ is the covariance matrix
of the single mode vacuum state given by diag(1/2, 1/2).
Similarly, on mixing the two modes using a balanced beam
splitter, the covariance matrix of the final TMSV state is
obtained as, \begin{equation} \label{covtmsvfinal}
V_{\text{TMSV}}=\frac{1}{2} \begin{pmatrix} \cosh (2r) & 0
&\sinh (2r) & 0\\ 0 & \cosh (2r) &0 & - \sinh (2r)\\ \sinh
(2r) & 0 &\cosh (2r) & 0\\ 0 & - \sinh (2r) &0 & \cosh
(2r)\\ \end{pmatrix}.  \end{equation} The mean vector
$\bm{d_\text{TMSV}}$ is pretty straightforward to calculate
and is given by $\bm{d_\text{TMSV}}= (0, 0, 0, 0)^T$\\

We now move on to the TMST state. As a prerequisite, we
first mention the density operator corresponding to a
thermal state of a single-mode system,
\begin{equation}\label{thermalfock}
\hat{\rho} = \sum_{n=0}^{\infty}
\frac{\langle n \rangle ^n}{(1+\langle n
\rangle)^{n+1}}|n\rangle \langle n|,
 \end{equation}
where $\langle n \rangle = 1/(\exp(\omega/k_B T)-1)$ is the
average
number of photons in the thermal state with $\omega$ being
the frequency of the mode. The covariance matrix of the
thermal state evaluates to
 \begin{equation}\label{intro:thermal}
 V_{\text{th}} =\frac{1}{2} \begin{pmatrix}
 2 \langle n \rangle+1 & 0 \\
 0 & 2 \langle n \rangle+1
\end{pmatrix}.
\end{equation}

To generate the TMST state, we follow the methodology
similar to that of generation of TMSV state. We first begin
with thermal states in both the modes $\rho_{\text{th}}
\otimes \rho_{\text{th}}$. Each of the modes is then
squeezed by an equal and opposite amount $r$. The
corresponding covariance matrix reads,
\begin{equation}
 \begin{aligned}
  V_{\text{initial}}&= [S_1(r)\oplus S_2(-r)][V_{\text{th}}
\oplus V_{\text{th}}] [S_1(r)\oplus S_2(-r)]^T, \\
 &= \left(
 \begin{array}{cccc}
  \kappa a^{-2 } & 0 & 0 & 0 \\
  0 & \kappa a^{2 } & 0 & 0 \\
  0 & 0 & \kappa a^{2 } & 0 \\
  0 & 0 & 0 & \kappa a^{-2 } \\
 \end{array}
 \right),
 \end{aligned}
 \label{covmatrixTMST}
\end{equation}
where $a=e^r$ and $\kappa=\langle n\rangle+1/2$.
The two modes are then combined using a balanced beam
splitter to obtain TMST state and the
resultant covariance matrix turns out to be
\begin{equation}
\label{covtmst}
 V_{\text{final}}= \begin{pmatrix}
  b & 0 &c & 0\\
  0 & b &0 & -c\\
  c & 0 &b & 0\\
  0 & -c &0 & b\\
 \end{pmatrix} 
\end{equation}
where \begin{equation}
 b= \frac{\left(a^4+1\right) \kappa}{2 a^2}, \quad c=
\frac{\left(a^4-1\right) \kappa}{2 a^2}.
\end{equation}
The mean vector $\bm{d_\text{TMST}}$ of the TMST state is
given by $\bm{d_\text{TMST}}= (0, 0, 0, 0)^T$\\

\subsubsection{ Wigner Characteristic function for Fock
state} The Wigner characteristic function for the fock state
$|n\rangle$ is given by,

\begin{equation}
\label{wignercharfock}
\chi(\tau, \sigma)_{|n\rangle}=\exp
\left(-\frac{\tau^2}{4}-\frac{\sigma^2}{4}\right)
L_n\left(\frac{\tau^2}{2}+\frac{\sigma^2}{2}\right),
\end{equation}

where $L_n$ denotes the Laguerre Polynomial.

\subsubsection{$U(n)$ Invariant Squeezing Criterion}
For an $n$ mode system with a covariance matrix $V$,
the state is said to be manifestly squeezed if any of the
diagonal elements
$V_{ii}<1/2$~\cite{simon-1994,arvind_pra_1995,ARVIND2002461}.
However, even if this condition is not met, squeezing may
still exist in other quadratures. This can be revealed by
virtue of passive $U(n)$ transformations which belong to
$K(n)$, the maximal compact subgroup of $Sp(2n,\,
\mathcal{R})$. $K(n)=SO(2n, \Re) \cap S p(2n, \Re)$ is
isomorphic to the unitary group $U(n)$ with elements
$K(X,Y)$ of the former mapping to elements $X-iY$ of
$U(n)$. Thus, under such a $U(n)$ invariant squeezing
approach, the search for squeezing is effectively conducted
across all the quadratures. In other words, a state is said
to be squeezed if and only if, $\left(K(X, Y) V K(X,
Y)^T\right)_{ii}<\frac{1}{2}$, for some $X-i Y \in U(n)$,
for some $i = 1,2, \dots ,2n$. This criteria remarkably
turns out to be equivalent to the minimum eigenvalue of the
covariance matrix being less than one-half as the necessary
and sufficient condition for a state to be squeezed. Again,
it is important to note that under such passive
transformations, the squeezed or unsqueezed status of a
state along with the minimum eigenvalue of the covariance
matrix remain unchanged. 

\subsection{Vaidman-Braunstein-Kimble quantum teleportation
protocol}
\label{bkprotocol}
We have followed the VBK protocol for the quantum
teleportation of an unknown input coherent state between two
distant parties, say Alice and Bob~\cite{Vaidman,bk-1998}.
The protocol requires the a priori existence of a two-mode
entangled resource state shared between the two parties. For
teleporting a specific input state, Alice mixes the input
mode with one of the modes of the entangled resource present
in his lab using a balanced beam splitter. The two output
modes are subjected to dual homodyne measurements, the
results of which are classically communicated to bob. Based
on the measurement results, Bob applies a specific
displacement operation on his part of the shared entangled
resource to recover the input state. Theoretically, Bob can
recover the input state exactly, however it requires the
resource state to be a maximally entangled state, such as the
TMSV state with infinite squeezing. In all practical
scenarios, the squeezing parameter is finite, and Bob can
only recover the input state approximately. The success of
the quantum teleportation protocol can be quantified via the
fidelity $F =\text{Tr} [\rho_{\text{in}}\rho_{\text{out}}]$,
where $\rho_{\text{in}}$ is the density operator of the
unknown input state and $\rho_{\text{out}}$ is the density
operator of the output (teleported) state. In Wigner
characteristic function formalism, the fidelity is written
as~\cite{Welsch-pra-2001}
\begin{equation}
\label{fidex1}
F =\frac{1}{2 \pi} \int d^2 \Lambda_2
\chi_{\text{in}}(\Lambda_2)
\chi_{\text{out}}(-\Lambda_2).
\end{equation}
It turns out that the Wigner characteristic function of the
output state can be expressed as a product of the Wigner
characteristic function of the input state and Wigner
characteristic function of the entangled resource 
state~\cite{Marian-pra-2006}:
\begin{equation}\label{teleported}
\chi_{\text{out}}(\tau_2,\sigma_2) =
\chi_{\text{in}}(\tau_2,\sigma_2) \chi_{A_1'
A_2'}(\tau_2,-\sigma_2,\tau_2,\sigma_2).
\end{equation}
It has been shown that in the absence of a shared entangled
resource state, the maximum achievable fidelity for
teleporting an input coherent state is
$1/2$~\cite{Braunstein-jmo-2000,Braunstein-pra-2001}. Hence,
the cases wherein fidelity exceeds $1/2$ are referred in the
article as successful quantum teleportation. 
\section*{Acknowledgement} A and C.K. acknowledge the
financial support from {\bf
DST/ICPS/QuST/Theme-1/2019/General} Project number {\sf
Q-68}. 

%

\end{document}